\documentclass
[superscriptaddress,secnumarabic,amssymb,amsmath,nobibnotes,aps,prd,nofootinbib,column,notitlepage,10pt]{revtex4}%
\usepackage{setspace}
\usepackage{color}
\usepackage{amsmath}
\usepackage{amsfonts}
\usepackage{verbatim}
\usepackage{amssymb}
\usepackage[colorlinks]{hyperref}
\usepackage{graphicx,bm}
\usepackage{amsmath}
\usepackage{amssymb}
\usepackage{epsfig}
\usepackage{epstopdf}
\usepackage{graphicx}
\usepackage{textgreek}
\usepackage[caption=false]{subfig}%
\usepackage{eurosym}
\usepackage{amssymb}
\usepackage{graphicx}
\usepackage{float}

\providecommand{\U}[1]{\protect\rule{.1in}{.1in}}

\newcommand{\be}{\begin{equation}}
\newcommand{\ee}{\end{equation}}

\newcommand{\mincir}{\raise
-3.truept\hbox{\rlap{\hbox{$\sim$}}\raise4.truept\hbox{$<$}\ }}
\newcommand{\magcir}{\raise
-3.truept\hbox{\rlap{\hbox{$\sim$}}\raise4.truept\hbox{$>$}\ }}

\ifx\pdfoutput\relax\let\pdfoutput=\undefined\fi
\newcount\msipdfoutput
\ifx\pdfoutput\undefined\else
\ifcase\pdfoutput\else
\ifx\paperwidth\undefined\else
\ifdim\paperheight=0pt\relax\else\pdfpageheight\paperheight\fi
\ifdim\paperwidth=0pt\relax\else\pdfpagewidth\paperwidth\fi
\fi\fi\fi
\begin{document}
\title{Hints Beyond $\Lambda$CDM from Barrow and Tsallis Holographic Dark Energy with GO cutoff}
\author{G. G. Luciano}
\email{giuseppegaetano.luciano@udl.cat}
\affiliation{Departamento de Qu\'{\i}mica, F\'{\i}sica y Ciencias Ambientales y del Suelo,
Escuela Polit\'{e}cnica Superior -- Lleida, Universidad de Lleida, Av. Jaume
II, 69, 25001 Lleida, Spain}
\author{A. Paliathanasis}
\email{anpaliat@phys.uoa.gr}
\affiliation{Institute of Systems Science, Durban University of Technology, Durban 4000, South Africa}
\affiliation{National Institute for Theoretical and Computational Sciences (NITheCS), South Africa}
\affiliation{Departamento de Matem\`{a}ticas, Universidad Cat\`{o}lica del Norte, Avda.
Angamos 0610, Casilla 1280 Antofagasta, Chile}
\author{G. Leon}
\email{genly.leon@ucn.cl}
\affiliation{Departamento de Matem\`{a}ticas, Universidad Cat\`{o}lica del Norte, Avda.
Angamos 0610, Casilla 1280 Antofagasta, Chile}
\affiliation{Institute of Systems Science, Durban University of Technology, Durban 4000, South Africa}
\author{A. Sheykhi}
\email{asheykhi@shirazu.ac.ir}
\affiliation{Department of Physics, College of Science, Shiraz University, Shiraz 71454, Iran}
\affiliation{Biruni Observatory, College of Science, Shiraz University, Shiraz 71454, Iran}
\author{M. Motaghi}
\affiliation{Department of Physics, College of Science, Shiraz
University, Shiraz 71454, Iran}

\begin{abstract}
Barrow and Tsallis Holographic Dark Energy (HDE) are two recent
extensions of the standard HDE framework, obtained by introducing
generalized entropy corrections through the Barrow and Tsallis
formalisms. In this work, we examine the cosmological consequences
of Barrow and Tsallis HDE implemented with the Granda-Oliveros
(GO) infrared (IR) cutoff. After deriving the modified Friedmann
equations within the thermodynamic-gravity conjecture, we study
the background evolution in both non-interacting and interacting
dark sector scenarios, emphasizing the role of the entropic
parameter in shaping late-time dynamics. We then confront the
model with state-of-the-art observations, including PantheonPlus
and Union3 Type Ia supernovae, Cosmic Chronometers and DESI DR2
BAO measurements. Using Bayesian MCMC methods, we constrain the
model parameters and compare the performance of BHDE with that of
$\Lambda$CDM. Our results show that BHDE is compatible with
current data and can exhibit a mild statistical preference over
the concordance model for certain dataset combinations. Overall,
the analysis underscores the relevance of generalized entropy
frameworks in late-time cosmology and identifies Barrow-Tsallis
holography with the GO cutoff as a competitive alternative to
$\Lambda$CDM.
\end{abstract}

\date{\today}
\maketitle

\section{Introduction}
\label{sec1}

The discovery of the accelerated expansion of the Universe at the
end of the twentieth century fundamentally reshaped modern
cosmology. Observations of distant type Ia supernovae revealed
that cosmic expansion is speeding
up~\cite{SupernovaSearchTeam:1998fmf,SupernovaCosmologyProject:1998vns},
implying the presence of an exotic energy component with strongly
negative pressure. This mysterious constituent, commonly referred
to as dark energy (DE), now represents the dominant contribution
to the cosmic energy budget, yet its origin and physical nature
remain largely
unknown~\cite{Bull:2015stt,Peebles:2002gy,Copeland:2006wr,Frieman:2008sn,Bamba:2012cp}.
The cosmological constant $\Lambda$ provides the simplest
description of DE and yields an excellent fit to a broad range of
cosmological observations. Nevertheless, the concordance
$\Lambda$CDM model faces persistent theoretical and observational
challenges~\cite{Perivolaropoulos:2021jda,CANTATA:2021asi,DiValentino:2021izs},
including the fine-tuning and coincidence problems, as well as
tensions in the measurements of $H_0$ and $\sigma_8$. These issues
have motivated the exploration of dynamical DE scenarios, in which
the equation of state (EoS) evolves with cosmic
time~\cite{Wetterich:1987fm,Tsujikawa:2013fta,Copeland:2006wr,Antoniadis:2006wq,Pan:2019brc,Najafi:2024qzm,Rezaei:2025vhb}.

A particularly intriguing class of dynamical DE models arises from
the holographic principle, which postulates that the number of
degrees of freedom within a spatial region scales with its
boundary area rather than its volume~\cite{tHooft:1993dmi}.
Applied to cosmology, this principle suggests that the DE density
may depend on an infrared (IR) length scale $L$ associated with
the cosmological horizon~\cite{Li:2004rb,Wang:2016och}, leading to
the holographic dark energy (HDE) density $\rho_{{DE}}\propto
L^{-2}$.

The cosmological behaviour of HDE depends crucially on the choice
of this cutoff. Several possibilities-such as the Hubble horizon,
particle horizon and future event horizon - have been proposed,
each leading to distinct physical and causal
properties~\cite{Wang:2016och}. To address limitations associated
with non-local cutoffs, Granda and Oliveros proposed a more
general, local IR scale depending on both the Hubble rate and its
time derivative~\cite{Granda:2008dk,Granda:2008tm}
\begin{equation}
\label{cutoff}
L = \left( \alpha H^2 + \beta \dot{H} \right)^{-1/2},
\end{equation}
where $\alpha$ and $\beta$ are dimensionless parameters. This
prescription avoids causality issues and naturally incorporates
the dynamical evolution of the cosmic expansion.

The standard HDE model is based on the Bekenstein-Hawking
entropy-area relation, which assumes the validity of semiclassical
gravity. However, this assumption may break down in non-extensive
or quantum-gravitational regimes. This possibility has prompted
the construction of several generalized HDE models, obtained by
replacing the conventional entropy with extended formulations such
as Tsallis~\cite{Tavayef:2018xwx,Saridakis:2018unr},
Renyi~\cite{Moradpour:2018ivi},
Sharma-Mittal~\cite{SayahianJahromi:2018irq}, and
Kaniadakis~\cite{Drepanou:2021jiv}
entropies~\cite{Nojiri:2021iko}. A further layer of physical
motivation for such extended approaches stems from the
thermodynamic foundations of gravitational
dynamics~\cite{Jacobson:1995ab,Padmanabhan:2003gd,Padmanabhan:2009vy,Paranjape:2006ca,Frolov:2002va,Wang:2001bf,Cai:2005ra,Cai:2006pa,Cai:2007bh,Cai:2007us}.
The well-established connection between the Friedmann equations
and the first law of thermodynamics at the apparent horizon
implies that any modification to the entropy-area relation induces
corresponding corrections to the cosmological background
equations. In this sense, generalized entropic frameworks predict
a modified Friedmann dynamics, offering a rich setting to explore
the interplay between horizon thermodynamics and cosmic
acceleration~\cite{Barrow:2020kug,Leon:2021wyx,Asghari:2021bqa,Luciano:2022pzg,Myrzakulov:2024jvg,Lymperis:2018iuz,Saridakis:2020lrg,Nojiri:2019skr,
Hernandez-Almada:2021rjs,Dheepika:2022sio,Jizba:2022icu,Lambiase:2023ryq,
Jizba:2024klq,Ebrahimi:2024zrk,Nojiri:2025gkq,
DAgostino:2019wko,DAgostino:2024sgm}.

Among the generalized entropic formulations introduced thus far,
Barrow entropy has recently attracted significant attention. It
introduces a quantum-gravitational deformation of the horizon
geometry, leading to a fractal-like structure. The ensuing entropy
takes the form~\cite{Barrow:2020tzx}
\begin{equation} \label{BE}
S_{\Delta} = \left( \frac{A}{A_{0}} \right)^{1+\frac{\Delta}{2}},
\end{equation}
where $A$ is the horizon area, $A_{0}=4G$ is the Planck area and
$\Delta$ is a dimensionless parameter characterizing the degree of
fractal deformation (we here adopt natural units). The classical
Bekenstein-Hawking entropy is recovered for $\Delta = 0$, while
$\Delta = 1$ represents a maximally deformed configuration.
Although Barrow's original proposal focused on $\Delta > 0$,
broader theoretical considerations suggest that negative values
may also be physically meaningful \cite{Tang, Dagotto:1989gp}. In
fact, recent observational analyses  indicate that mildly negative
$\Delta$ could actually be favoured by cosmological
data~\cite{Luciano:2025elo,Luciano:2025hjn}. Moreover, despite
arising from conceptually distinct frameworks, the Barrow entropy
\eqref{BE} possesses the same functional form as the exponent
$\epsilon$ in Tsallis entropy~\cite{Tsallis:2013} under the
correspondence $\Delta \rightarrow 2(\epsilon-1)$. Consequently,
although the subsequent analysis is developed within the Barrow
framework, this formal equivalence ensures that the resulting
cosmological dynamics and constraints can be straightforwardly
extended to the Tsallis case as well.

The generalized entropy-area relation~\eqref{BE} has been shown to
naturally modify the HDE density, leading
to~\cite{Saridakis:2020zol}
\begin{equation}
\label{BHDE}
\rho_{DE} = C L^{\Delta - 2}\,,
\end{equation}
where $C$ has dimensions $[L]^{-2-\Delta}$. In the limit $\Delta =
0$, the standard HDE form is recovered with $C = 3 c^2 M_p^2$,
where $c$ is a dimensionless constant and $M_p^2 = (8 \pi G)^{-1}$
the reduced Planck mass. The resulting Barrow Holographic Dark
Energy (BHDE) model has been extensively studied in various
cosmological contexts. The inclusion of the Barrow exponent
$\Delta$ enhances the compatibility of HDE with observational
data, alleviating tensions related to the dark energy EoS and the
Hubble parameter at low redshifts. It also enriches the
dark-sector dynamics, enabling a smoother transition between
cosmological epochs and a more flexible description of late-time
acceleration~\cite{Anagnostopoulos:2020ctz,Dabrowski:2020atl,Srivastava:2020cyk,Adhikary:2021xym,Nojiri:2021jxf,Luciano:2022hhy,Sheykhi:2022fus,Luciano:2022ffn,Luciano:2023wtx,Li:2024bwr,Luciano:2025hjn,Basilakos:2023seo,Luciano:2025elo}.

Within the BHDE framework, several IR cutoffs have been
considered~\cite{Saridakis:2020zol,Srivastava:2020cyk,Adhikary:2021xym,Anagnostopoulos:2020ctz,Dabrowski:2020atl}.
Among these developments, the incorporation of the GO cutoff has
been recently
investigated~\cite{Oliveros:2022biu,Motaghi:2024rag}, showing that
the combination of Barrow entropy with the dynamical GO scale
yields a causally consistent and observationally viable model.
This scenario predicts a delayed deceleration/acceleration
transition and remains stable for specific parameter ranges,
providing a cosmic evolution compatible with the latest data in
both flat and non-flat geometries.
Further studies have been recently conducted in this direction~\cite{Mahmoudifard:2025PDU,Mahmoudifard:2024EPJC,Yarahmadi:2024MNRAS},
where the BHDE model with the GO
IR cutoff was explored from different perspectives.
Specifically, Ref.~\cite{Mahmoudifard:2025PDU} investigated the theoretical
consistency of Barrow cosmology with the GO cutoff and confirmed the
existence of a viable late-time accelerated phase.
In Ref.~\cite{Mahmoudifard:2024EPJC}, observational data were employed
to constrain the model parameters and neutrino masses, showing that
the BHDE-GO scenario remains compatible with current cosmological
bounds.
Moreover, Ref.~\cite{Yarahmadi:2024MNRAS} demonstrated that the same
framework can significantly alleviate the Hubble tension.
We will show that our results are in qualitative agreement with these works regarding
the late-time acceleration of the BHDE--GO model, while
extending the analysis by providing a complementary investigation of
the cosmological dynamics, thereby further supporting the robustness
and observational viability of the BHDE
scenario.

Starting from the above premises, in this work we confront the
BHDE model constructed with the GO IR cutoff with a broad set of
current cosmological observations. In particular, we employ the
Supernova PantheonPlus (PP) and Union3 (U3) compilations, the
Observational Hubble Data (OHD) and the Baryon Acoustic
Oscillation (BAO) measurements to constrain the model parameters
and assess its phenomenological viability. The statistical
performance of the BHDE scenario is then compared with that of the
concordance $\Lambda$CDM model. Our analysis shows that, for
certain data combinations, the extended BHDE framework provides a
better fit to the observations and exhibits a weak but
non-negligible statistical preference over $\Lambda$CDM, while
remaining fully consistent with the standard cosmological
constraints.

The paper is structured as follows. In the next section, we review
the derivation of the modified Friedmann equations in the context
of Barrow cosmology, outlining the main theoretical ingredients
relevant to the BHDE framework. In Sec.~\ref{sec3}, we focus on
the BHDE model  with the GO cutoff, investigating both the
non-interacting and interacting cases between the dark components.
Observational analysis is conducted in Sec. \ref{sec4}, while
conclusions and outlook are summarized in Sec. \ref{sec5}.
\section{Modified Friedmann equations from Barrow entropy}
\label{sec2}

We begin by revisiting the derivation of the modified Friedmann
equations emerging from the Barrow entropy formalism. Within the
thermodynamic approach to gravity, the cosmological dynamics can
be obtained by applying the first law of thermodynamics to the
apparent horizon, taking into account the entropy deformation
proposed by Barrow~\cite{Sheykhi:2021fwh}.

Our analysis is performed within a spatially flat
Friedmann-Lema\^{\i}tre-Robertson-Walker (FLRW) background,
described by the metric \be ds^2 =h_{\alpha\beta}dx^\alpha
dx^\beta + \tilde r^2\left(d\theta^2 + \sin^2\theta\,
d\phi^2\right)\, , \label{FLRW_metric} \ee where $\tilde r =
a(t)\,r$, $x^0 = t$, $x^1 = r$, $h_{\alpha\beta} =
\mathrm{diag}(-1,a^2)$, and $a(t)$ denotes the time-dependent
scale factor. We consider the apparent horizon as the effective
boundary of the Universe, with radius $\tilde r_{A} = 1/H$, where
the Hubble parameter $H = \dot{a}/a$ characterizes the rate of
cosmic expansion, and the overdot denotes differentiation with
respect to cosmic time \cite{Frolov:2002va,Cai:2005ra,Cai:2009qf}.
This choice is compatible with both the first and second laws of
thermodynamics.

Since the work done on the system is proportional to the change in
volume ($dV$), the first law applied at the apparent horizon can
be expressed as
\begin{equation}
\label{FL}
dE=TdS+\mathcal{W}dV\,,
\end{equation}
where the horizon temperature takes the Hawking-like expression~\cite{Hawking:1975vcx}
\begin{equation}
\label{temp}
 T=-\frac{1}{2 \pi \tilde r_A}\left(1-\frac{\dot {\tilde
r}_A}{2H\tilde r_A}\right),
\end{equation}
in analogy with black hole
thermodynamics~\cite{Cai:2009qf,Padmanabhan:2009vy}. Furthermore,
the work density $\mathcal{W}$, which arises from variations of
the apparent horizon radius, is given by $\mathcal{W} =
-\frac{1}{2}\,\text{Tr}(T^{\mu\nu}) = (\rho-p)$/2, where
$T_{\mu\nu} = (\rho + p)\, u_{\mu} u_{\nu} + p\, g_{\mu\nu}$ is
the energy-momentum tensor of the perfect fluid that fills the
Universe. Here, $\rho$ denotes the energy density, $p$ is the
isotropic pressure, $u^{\mu}$ is the four-velocity of a comoving
observer satisfying $u^{\mu}u_{\mu} = -1$ and $g_{\mu\nu}$
represents the spacetime metric. The trace $\text{Tr}(T^{\mu\nu})$
is evaluated with respect to the induced metric on the $(t,r)$
submanifold of the FLRW geometry.

By differentiating the total matter and energy $E = \rho V$ within a
three-dimensional sphere of radius $\tilde r_A$, and using the continuity
equation $\dot{\rho} + 3H(\rho + p) = 0$,
we obtain
\begin{equation}
\label{dE}
dE = 4\pi \tilde r_A^2 \rho \, d\tilde r_A - 4\pi H \tilde r_A^3 (\rho + p)\, dt\,.
\end{equation}
On the other hand, differentiating the Barrow entropy expression \eqref{BE}
yields
\begin{equation}
\label{dS}
dS = \left(2 + \Delta\right) \left(\frac{4\pi}{A_0}\right)^{1 + \Delta/2} \, \tilde r_A^{\,1 + \Delta} \,
\dot{\tilde r}_A \, dt\,.
\end{equation}
By substituting Eqs. \eqref{temp}, \eqref{dE} and \eqref{dS} into the first law of thermodynamics \eqref{FL} and integrating, we finally obtain \cite{Motaghi:2024rag}
\begin{equation}
\label{MFE}
    H^{2-\Delta}=\frac{8\pi G_{\rm eff}}{3}\,\rho\,,
\end{equation}
where we have expressed the horizon radius in terms the Hubble rate. Furthermore, we have defined the effective gravitational constant
\begin{equation}
\label{effG} G_{\rm
eff}=\dfrac{A_0}{4}\left(\dfrac{2-\Delta}{2+\Delta}\right)\left(\dfrac{A_0}{4\pi}\right)^{\Delta/2}\,.
\end{equation}
The relation \eqref{MFE} provides the modified Friedmann equation
obtained from Barrow entropy. It is worth noting that in this
formulation, the corrections arising from the modified entropy are
incorporated through a redefinition of the gravitational constant.
A conceptually different interpretation is provided in Ref.
\cite{Saridakis:2020lrg}, where the $\Delta$-dependent terms are
treated as an additional contribution to the dark energy sector.
Moreover, it is straightforward to verify that in the limit
$\Delta \to 0$, the effective gravitational constant reduces to
the standard value, $G_{\rm eff} = G$, thereby recovering the
conventional Friedmann dynamics.
\section{BHDE with GO cutoff} \label{sec3} We consider the BHDE model during an epoch in which
the cosmic fluid contains both (pressureless) dark matter (DM) and
DE. For later convenience, we rewrite Eq.~\eqref{MFE} in the form
\begin{equation}
\label{MFE2}
    H^{2-\Delta}=\frac{1}{3M^2_{\rm eff}}\left(\rho_m+\rho_{DE}\right),
\end{equation}
where $\rho_m$ is the matter energy density and $M^2_{\rm
eff}\equiv(8\pi G_{\rm eff})^{-1}$.

By combining the BHDE density \eqref{BHDE} with the GO cutoff \eqref{cutoff}, the energy density can be written as
\begin{equation}
    \rho_{DE} = 3 M_{\rm eff}^2 \left( \alpha H^2 + \beta \dot{H} \right)^{1-\Delta/2} \,,
    \label{NEWBHDE}
\end{equation}
where we have fixed the constant $C=3 M_{\rm eff}^2$ (equivalent
to setting  $c^2 = 1$ in the standard normalization). In the limit
$\Delta \rightarrow 0$, the results of
\cite{Granda:2008dk,Granda:2008tm} are fully recovered, ensuring
consistency with the standard GO HDE scenario.
\subsection{Non-interacting (NI) case}
We start by assuming that the dark sectors are conserved independently.
In this case, the corresponding conservation equations read
\begin{eqnarray}
&\dot\rho_{DE}+3H\rho_{DE}\left(1+\omega_{DE}\right)=0\,,&\\[2mm]
&\dot\rho_m+3H\rho_m=0\,,&
\end{eqnarray}
where $\omega_{DE} \equiv p_{DE}/\rho_{DE}$ is the EoS parameter
of the DE component.

It is now convenient to introduce the effective critical energy
density $\rho_{cr}=3M_{\rm eff}^2H^{2-\Delta}$. In turn, the
fractional energy densities of DM and DE are
\begin{equation}
\label{frac}
    \Omega_{m}\equiv \frac{\rho_m}{\rho_{cr}}=
    \frac{\rho_{m}}{3M^2_{\rm eff}H^{2-\Delta}}\,,\qquad \,\,
    \Omega_{DE}\equiv \frac{\rho_{DE}}{\rho_{cr}}=\frac{\rho_{DE}%
}{3M_{\rm eff}^2H^{2-\Delta}}\,,%
\end{equation}
which enables us to rewrite the modified Friedmann equation \eqref{MFE2} in the form $\Omega_m+\Omega_{DE}=1$.

To explore the dynamics of the DE model throughout the history of
the Universe, one must analyze the evolution of the DE density
parameter $\Omega_{DE}$. Using Eqs.~\eqref{MFE2} and \eqref{frac}
and some straightforward algebra, it can be shown that
\cite{Motaghi:2024rag}
\begin{equation}
\label{OMDE}
\frac{d\Omega_{DE}}{dz}=\frac{3\left(1-\Omega_{DE}\right)}{\left(1+z\right)}
\left[\frac{\left(\Delta-2\right)}{3\beta}\left(\Omega_{DE}^{2/\left(2-\Delta\right)}-\alpha\right)-1\right],
\end{equation}
where $z=1/a-1$ denotes the redshift variable (we set the
present-day value of $a$ to unity). The evolution of $\Omega_{DE}$
has been studied in \cite{Motaghi:2024rag}. Interestingly, the
results indicate that for $z \gtrsim 0.6$, $\Omega_{DE}$ decreases
with increasing values of $\Delta$, whereas at lower redshifts the
dependence on $\Delta$ becomes negligible.
The EoS parameter and the deceleration parameter
are given by \cite{Motaghi:2024rag}
\begin{align}\label{wd}
w_{DE}=\dfrac{1}{\Omega_{DE}}\left[ \dfrac{(
\Delta-2)}{3\beta}(\Omega_{DE}^{2/(2-\Delta)}-\alpha)-1\right],
\end{align}
\begin{align}\label{q1}
q=-1-\dfrac{\dot{H}}{H^2}=-1-
\dfrac{\Omega_{DE}^{2/(2-\Delta)}-\alpha}{\beta}.
\end{align}

\subsection{Interacting (I) case}
Next, we consider a FRW Universe composed of BHDE and DM, allowing
for energy exchange between them. Since the fundamental nature of
both dark components remains unknown, there is sufficient freedom
to permit an interaction between these components
\cite{Wetterich:1987fm}. Moreover, in \cite{Pereira:2008at} the
interaction between DM and DE was thoroughly examined from a
thermodynamic perspective.

In this scenario, the semi-conservation equations accounting for
the interaction take the form
\begin{eqnarray}
&\dot\rho_{DE}+3H\rho_{DE}\left(1+\omega_{DE}\right)=-Q\,,&\\[2mm]
&\dot\rho_m+3H\rho_m=Q\,,&
\end{eqnarray}
where \(Q\) parameterises the energy exchange between DE and DM. A
variety of phenomenological ans\"atze for \(Q\) have been
considered in the literature \cite{Wang:2016lxa}. Following
\cite{Pavon:2005yx,Motaghi:2024rag}, we set
\begin{equation}
\label{eq:Q_choice}
Q=3\gamma\hspace{0.3mm} H\left(1+r\right)\rho_{DE}\,,
\end{equation}
where $\gamma=b^2>0$ is a coupling constant and $r\equiv
\rho_m/\rho_{DE}$ \cite{Pavon:2005yx}. Since $Q>0$ enters the DE
continuity equation with a minus sign, the interaction term
effectively describes an energy transfer from DE to DM.

The form \eqref{eq:Q_choice} is largely used for several reasons.
First, the factor \(H\) guarantees the correct dimensionality for
a transfer rate and ties the strength of the interaction to the
global expansion, while the additional term
\(\left(1+r\right)\rho_{DE}=\rho_{DE}+\rho_{m}\) ensures that the
coupling is effectively proportional to the total energy density
of the dark sector, thereby allowing the interaction to remain
dynamically relevant both at early and late times.  Moreover, the
model retains phenomenological flexibility while involving only
one additional dimensionless parameter, making it straightforward
to implement in cosmological analyses and to constrain
observationally \cite{Pavon:2005yx}.

Following calculations similar to those developed for the non-interacting model, it can be shown that the DE density obeys the generalized dynamics \cite{Motaghi:2024rag}
\begin{equation}
\label{intmo}
\frac{d\Omega_{DE}}{dz}=\frac{3}{\left(1+z\right)}\left\{\left(1-\Omega_{DE}\right)\left[
\frac{
\left(\Delta-2\right)}{3\beta}\left(\Omega_{DE}^{2/\left(2-\Delta\right)}-\alpha\right)-1\right]+\gamma\right\}\,,
\end{equation}
which straightforwardly recovers Eq. \eqref{OMDE} for $\gamma=0$.
Compared to the previous model, the effects of the interaction are
manifested in a modified evolution of $\Omega_{DE}$, which, for
each $z$, is observed to decrease as $\gamma$ decreases.
It also
shapes the evolution of the EoS parameter $w_{DE}$, which
decreases as $\gamma$ increases, reflecting the distinct ways in
which the coupling influences each quantity. Notably, for
sufficiently small values of $\gamma$, $w_{DE}$ remains in the
phantom regime at the present epoch \cite{Motaghi:2024rag}.

For later convenience, we introduce a rescaling of the variables and rewrite Eq. (\ref{intmo}) in the form
\begin{equation}
\frac{d\Omega_{DE}}{dz}=\frac{3}{\left(1+z\right)}\left\{\left(1-\Omega_{DE}\right)\left[
\left(B\Omega_{DE}^{2/\left(2-\Delta\right)}-A\right)-1\right]+\gamma\right\}\,,
\label{master2}
\end{equation}
where $B=\frac{\Delta-2}{3 \beta}$ and $A=\alpha B$. The EoS
parameter and the deceleration parameter for the interacting case,
are still given by Eqs. (\ref{wd}) and (\ref{q1})
\cite{Motaghi:2024rag}, while in this case the evolutionary form
of $\Omega_{DE}$ is governed by (\ref{master2}).

\section{Observational Data Analysis}\label{sec4}
In this section, we utilize different sets of observational data
to place constraints on the free parameters of the BHDE model with
the GO IR cutoff. By confronting the theoretical predictions with
measurements from cosmological probes, we aim to assess the
viability of the model and quantify the allowed parameter space.
\subsection{Observational Data}
Let us first describe the specific datasets employed in this
analysis:
\begin{itemize}
\item \textbf{PantheonPlus Supernovae (PP):}
The PantheonPlus compilation includes a total of 1701 light curves
corresponding to 1550 spectroscopically confirmed Type Ia
supernovae. It provides observational measurements of the distance
modulus $\mu^{\mathrm{obs}}$ across the redshift interval $10^{-3}
< z < 2.27$~\cite{pan}. The theoretical prediction for the
distance modulus is given by
\begin{equation}
\mu^{\mathrm{th}} = 5 \log_{10} D_{L} + 25,
\end{equation}
where, under the assumption of a spatially flat FLRW metric, the luminosity distance is expressed as
\begin{equation}
D_{L}(z) = (1 + z) \int_{0}^{z} \frac{dz'}{H(z')}.
\end{equation}
In this work, we employ the PantheonPlus sample without incorporating the SH0ES Cepheid calibration.

\item \textbf{Union3 Supernovae (U3):}
The Union3 catalogue represents the most up-to-date supernova
dataset, containing 2087 events within the same redshift interval
as the PP compilation~\cite{union}. Of these, 1363 supernovae are
common to both the Union3 and PantheonPlus catalogues.

\item \textbf{Observational Hubble Data (OHD):}
We utilize direct measurements of the Hubble parameter $H(z)$
derived from the Cosmic Chronometers (CC) method. These
observations are model-independent, relying solely on the
differential aging of passively evolving galaxies with
synchronized stellar populations and homogeneous cosmic
evolution~\cite{co01}. Specifically, we employ 34 $H(z)$ data
points spanning the redshift range $0.09 \leq z \leq 1.965$, as
reported in~\cite{cc1} and the three new measurements
from the analysis of the DESI DR1 data \cite{Loubser:2025fzl}.

\item \textbf{Baryon Acoustic Oscillations (BAO):}
We incorporate the latest BAO measurements from the Dark Energy Spectroscopic Instrument (DESI) Data Release 2 (DR2)~\cite{des4,des5,des6}.
This dataset provides constraints on several cosmological distance ratios evaluated at seven distinct redshifts, namely:
\begin{equation}
\frac{D_{M}}{r_{\mathrm{drag}}} = \frac{D_{L}}{(1+z)\,r_{\mathrm{drag}}}, \quad\,\,\,\,
\frac{D_{V}}{r_{\mathrm{drag}}} = \frac{(z D_{H} D_{M}^{2})^{1/3}}{r_{\mathrm{drag}}}, \quad\,\,\,\,
\frac{D_{H}}{r_{\mathrm{drag}}} = \frac{1}{r_{\mathrm{drag}} H(z)},
\end{equation}
where $D_{L}$ is the luminosity distance and $r_{{drag}}$ denotes the comoving sound horizon at the baryon-drag epoch.
In our analysis, $r_{{drag}}$ is treated as a free parameter.
\end{itemize}
\subsection{Methodology}
To perform the statistical inference, we employ the Bayesian
analysis framework
\textsc{Cobaya}\footnote{\url{https://cobaya.readthedocs.io/}}~\cite{cob1,cob2},
using a customized theoretical module coupled with its built-in
Markov Chain Monte Carlo (MCMC) sampler~\cite{mcmc1,mcmc2}. The
resulting MCMC chains are post-processed and visualized using the
\textsc{GetDist}
package\footnote{\url{https://getdist.readthedocs.io/}}~\cite{getd}.

We analyze two distinct combinations of cosmological datasets, namely $\mathrm{PP\&OHD\&BAO}$ and $\mathrm{U3\&OHD\&BAO}$. For each dataset configuration, the model
parameters are estimated by maximizing the likelihood function, defined as
$\mathcal{L}_{\max} = \exp\!\left(-\tfrac{1}{2}\chi_{\min}^{2}\right)$,
where the minimum chi-square value corresponds to the sum of the individual
contributions from each dataset.
For comparison purposes, the same statistical analysis is also carried out
for the reference $\Lambda$CDM model.

Since the BHDE framework and the standard
$\Lambda$CDM cosmology contain a different number of free parameters,
a fair model comparison requires an information-based criterion.
We therefore employ the Akaike Information Criterion (AIC)~\cite{AIC}, defined as
\begin{equation}
\mathrm{AIC} = \chi_{\min}^{2} + 2\kappa,
\end{equation}
where $\kappa$ denotes the total number of independent parameters in the model.
Lower AIC values correspond to a better balance between model complexity and
goodness of fit.

To quantify the statistical preference between the two models,
we compute the relative difference
\begin{equation}
\Delta \mathrm{AIC} = \mathrm{AIC}_{\mathrm{BHDE}} - \mathrm{AIC}_{\Lambda\mathrm{CDM}}.
\end{equation}
Because the BHDE framework includes more free parameters than the standard
$\Lambda$CDM model (see Tab. \ref{prior}), the correction term in the AIC expression varies according
to whether the BHDE model is interacting or not.
In the non–interacting configuration, the additional degrees of freedom are
$\{\Delta, \alpha, \beta\}$, corresponding to $\Delta\kappa = 3$. Therefore
\begin{equation}
\Delta \mathrm{AIC}_{\mathrm{NI}}
= \chi_{\min}^{2}(\mathrm{BHDE_{NI}})
- \chi_{\min}^{2}(\Lambda\mathrm{CDM}) + 6\,.
\end{equation}
On the other hand, in the interacting case an extra parameter associated with the coupling
strength $\gamma$ is introduced, giving $\Delta\kappa = 4$ and
\begin{equation}
\Delta \mathrm{AIC}_{\mathrm{I}}
= \chi_{\min}^{2}(\mathrm{BHDE_{I}})
- \chi_{\min}^{2}(\Lambda\mathrm{CDM}) + 8\,.
\end{equation}

Following Akaike's classification, models with $|\Delta
\mathrm{AIC}| < 2$ are statistically indistinguishable. Values in
the range $2 < |\Delta \mathrm{AIC}| < 6$ indicate weak evidence
in favor of the model with the lower AIC, while $6 < |\Delta
\mathrm{AIC}| < 10$ corresponds to strong evidence. Finally, a
difference exceeding $10$ denotes decisive support for the model
with the smaller AIC value.

To ensure consistency in the comparison and to mitigate any potential
systematic bias, the Hubble expansion rate for the $\Lambda$CDM model
has been computed numerically using the same procedure applied to the BHDE framework.
\begin{table}[t] \centering
\begin{tabular}
[c]{ccc}\hline\hline
\textbf{Priors} & \textbf{BHDE} & $\Lambda$\textbf{CDM}\\\hline
$\mathbf{H}_{0}$ & $\left[  60,80\right]  $ & $\left[  60,80\right]  $\\
$\mathbf{\Omega}_{m0}$ & $\left[  0.01,0.5\right]  $ & $\left[
0.01,0.5\right]  $\\
$\mathbf{r}_{drag}$ & $\left[  130,160\right]  $ & $\left[  130,160\right]
$\\
$\Delta$ & $\left[  -1,1\right]  $ & $-$\\
$A$ & $\left[ -5,0\right]  $ & $-$\\
$\mathbf{B}$ & $\left[  -5,0\right]  $ & $-$\\
$\gamma$ & $\left[  0,1\right]  $ & $-$\\\hline\hline
\end{tabular}
\caption{Priors adopted for the free parameters of the $\Lambda$CDM and BHDE models.
For $\Lambda$CDM, the parameter set is $\{ H_0, \Omega_{m0}, r_{\rm drag} \}$,
while for BHDE it includes $\{ H_0, \Omega_{m0}, r_{\rm drag}, \Delta, \alpha, \beta \}$ for the non-interacting model and $\{ H_0, \Omega_{m0}, r_{\rm drag}, \Delta, \alpha, \beta, \gamma \}$ for the interacting model.
}
\label{prior}%
\end{table}%

\subsection{Results}

We now present the results obtained from the analysis of the MCMC
chains for the combined datasets PP\&OHD\&BAO and U3\&OHD\&BAO,
for the two models under consideration, namely, the
non-interacting (NI) BHDE and the interacting (I) BHDE.
\subsubsection{Non-interacting case}
The contour plots of the confidence
regions for the best-fit parameters of the non-interacting BHDE model are displayed in Fig. \ref{fig1}.
For the PP\&OHD\&BAO dataset, we obtain the posterior estimates
$H_{0}=67.9_{-1.7}^{+1.7}\,\text{km\,s}^{-1}\,\text{Mpc}^{-1}$,
$\Omega_{m0}=0.358_{-0.02}^{+0.16}$,
$r_{drag}=147.5_{-3.5}^{+3.5}\,\text{Mpc}$,
$\Delta=-0.041_{-0.34}^{+0.01}$,
$\alpha = 1.58_{-0.55}^{+0.23}$,
and $\beta = 1.49_{-0.93}^{+0.49}$ (see Tab. \ref{bestfit}).
This model provides a slightly better fit to the data than $\Lambda$CDM, as indicated by
$\chi_{\min}^{2}-\chi^{2}_{\Lambda,\min} = -4.27$.
However, once the different numbers of degrees of freedom are taken into account, the Akaike information criterion shows that the two models are statistically equivalent, with
$\Delta\mathrm{AIC}_{\mathrm{NI}} = +1.73$.

On the other hand, the introduction of the U3 catalogue, and for the dataset
U3\&OHD\&BAO we find the posterior variables $H_{0}=66.2_{-1.7}^{+1.7}$,
$\Omega_{m0}=0.462_{-0.044}^{+0.044}$,~$r_{drag}=148.0_{-3.5}^{+3.5}$,
$\Delta=-0.234_{-0.093}^{+0.082}$,~$\alpha>14.6$ and $\beta=23_{-7}^{+10}$  (see Tab. \ref{bestfit}). Unlike the previous case, the data now show a weak preference for the BHDE model, with
$\chi_{\min}^{2}-\chi^{2}_{\Lambda,\min} = -9.63$ and $\Delta\mathrm{AIC}_{\mathrm{NI}}= -3.63$.

It is worth commenting on the fact that the best-fit value of
$\Delta$ turns out to be negative in both datasets. In the
original Barrow proposal, positive values $\Delta>0$ were
introduced to model ``sphereflake''-type deformations of the
horizon, leading to a surface with enhanced roughness. However,
negative values of $\Delta$ correspond to ``porous'' or ``spongy''
geometries, which are also well motivated from the viewpoint of
fractal geometry. Indeed, many classical fractal constructions
possess effective anomalous dimensions below the topological one
\cite{Tang,Xu}. Moreover, from the quantum field theory
perspective, renormalization-group arguments allow for negative
anomalous dimensions in several systems \cite{Dagotto:1989gp}.
Even more significantly, negative logarithmic corrections to the
Bekenstein--Hawking entropy - such as those obtained from the
Cardy formula \cite{Carlip:2000nv} - indicate that $\Delta$ may
naturally take negative values.

Furthermore, we remark on the parameters $\alpha$ and $\beta$.
According to \cite{Manoharan:2024thb}, these parameters are
expected to lie in the ranges $\alpha \in [0.7,1.0]$ and $\beta
\in [0.3,0.9]$. For the PP\&OHD\&BAO dataset, our best-fit values
are marginally consistent with these ranges within $1\sigma$
uncertainties. For the U3\&OHD\&BAO dataset, the estimates of
$\alpha$ and $\beta$ slightly exceed the expected ranges. These
discrepancies can be attributed to the fact that the study of
\cite{Manoharan:2024thb} focuses on redshifts $z \gtrsim 2$,
whereas our datasets primarily probe lower-redshift regimes, in
which the effective dynamics of the BHDE model may differ.
\subsubsection{Interacting case}
For the second model in our analysis, where interaction is
included, the contour plots of the confidence regions for the
best-fit parameters are shown in Fig. \ref{fig2}. We find for the
PP\&OHD\&BAO dataset the posterior parameters
$H_{0}=67.9_{-1.7}^{+1.7}\,\text{km\,s}^{-1}\,\text{Mpc}^{-1}$,
$\Omega_{m0}=0.418_{-0.029}^{+0.079}$,
$r_{drag}=147.5_{-3.4}^{+3.4}\,\text{Mpc}$,
$\Delta=-0.197_{-0.16}^{+0.06}$, $\alpha=1.20_{-0.56}^{+0.25}$,
$\beta=1.31_{-0.78}^{+0.29}$, and $\gamma<0.086$ (see Tab.
\ref{bestfit}). In comparison with the $\Lambda$CDM model, the
statistical parameter is
$\chi_{\min}^{2}-\chi_{\Lambda,\min}^{2}=-4.27$, indicating that
the BHDE models with and without interaction fit the data
similarly. However, due to the different number of degrees of
freedom, the data provide a weak preference for $\Lambda$CDM,
since $\Delta \mathrm{AIC_{\mathrm{I}}}=+3.73$ in this case.
Furthermore, the non-interacting model, which corresponds to
$\gamma=0$, lies within the $1\sigma$ posterior range.

For the dataset U3\&OHD\&BAO the analysis of the MCMC chains give
$H_{0}=65.8_{-1.7}^{+1.7}$, $\Omega_{m0}=0.480_{-0.044}^{+0.039}$%
,~$r_{drag}=148.1_{-3.5}^{+3.5}$, $\Delta=-0.260_{-0.083}^{+0.083}$%
,~$\alpha=17.0_{-5.0}^{+12}$, $\beta=30_{-8}^{+20}$ and $\gamma<0.127$. The
two BHDE models fit the data with the same way, $\chi_{\min}^{2}\mathbf{-}%
\chi_{\Lambda\min}^{2}=-9.67$. Nevertheless, due to the different degrees of
freedom, the interacting model is statistical equivalent with the $\Lambda
$CDM, i.e. $AIC-AIC_{\Lambda}=-1.67$%

Analogously to the discussion presented above, the negative values
of $\Delta$ are preferred. As for the parameters $\alpha$ and
$\beta$, the estimates obtained from the PP\&OHD\&BAO dataset are
broadly consistent with the ranges reported in
\cite{Manoharan:2024thb}, while those from the U3\&OHD\&BAO
dataset show a mild tension, particularly for $\beta$. Finally,
the interaction parameter $\gamma$ is tightly constrained in both
datasets, with upper limits compatible with typical values
considered in the literature \cite{Xia:2016vnp}. These results
indicate that, while a non-zero interaction cannot be excluded,
the data are also consistent with the non-interacting limit,
showing that current observations place only weak constraints on
the strength of the dark-sector coupling.

\begin{figure}[t]
\centering\includegraphics[width=0.9\textwidth]{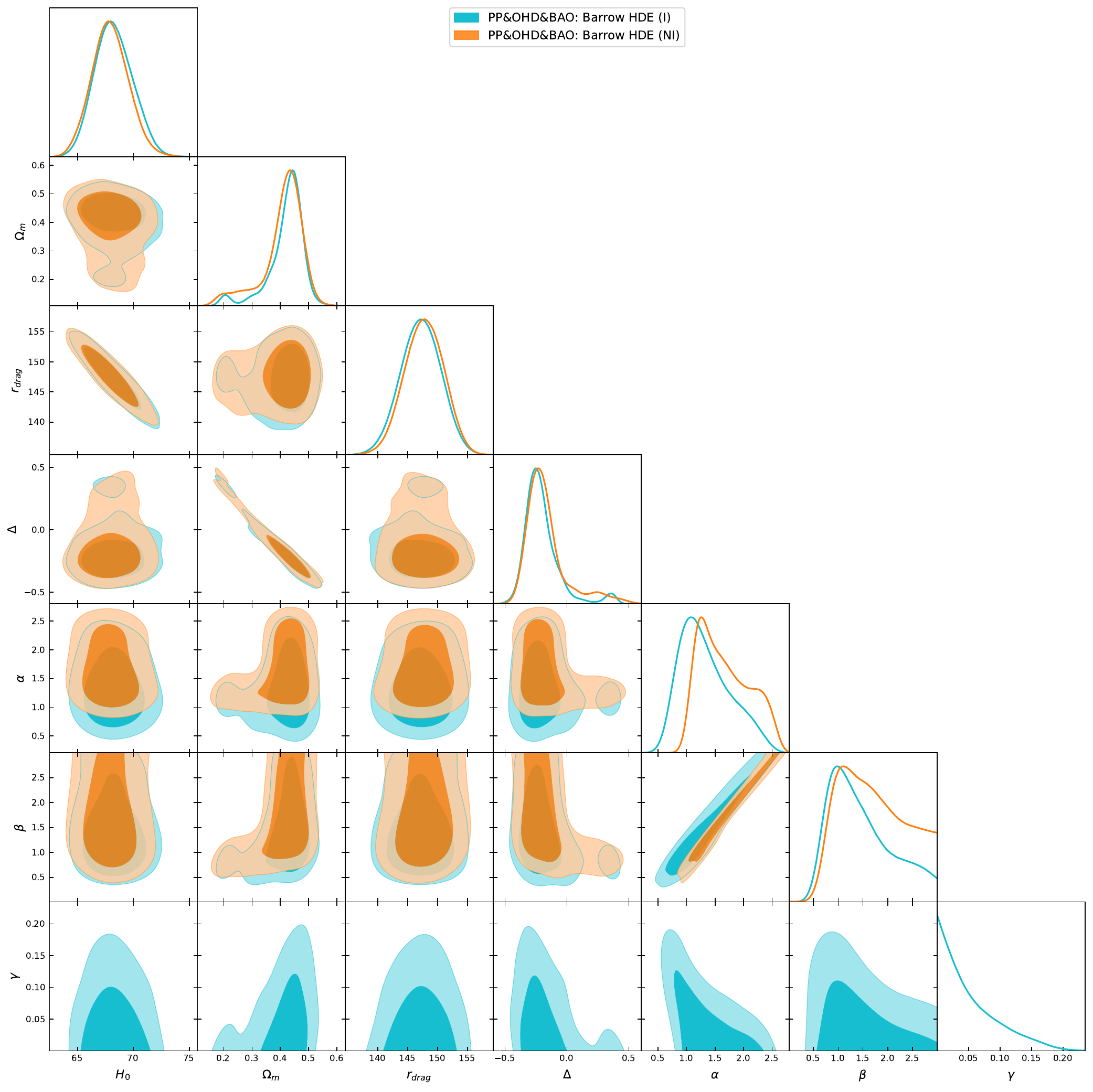}\caption{Confidence
space for the posterior parameters for the non-interacting BHDE model.}%
\label{fig1}%
\end{figure}

\begin{figure}[t]
\centering\includegraphics[width=0.9\textwidth]{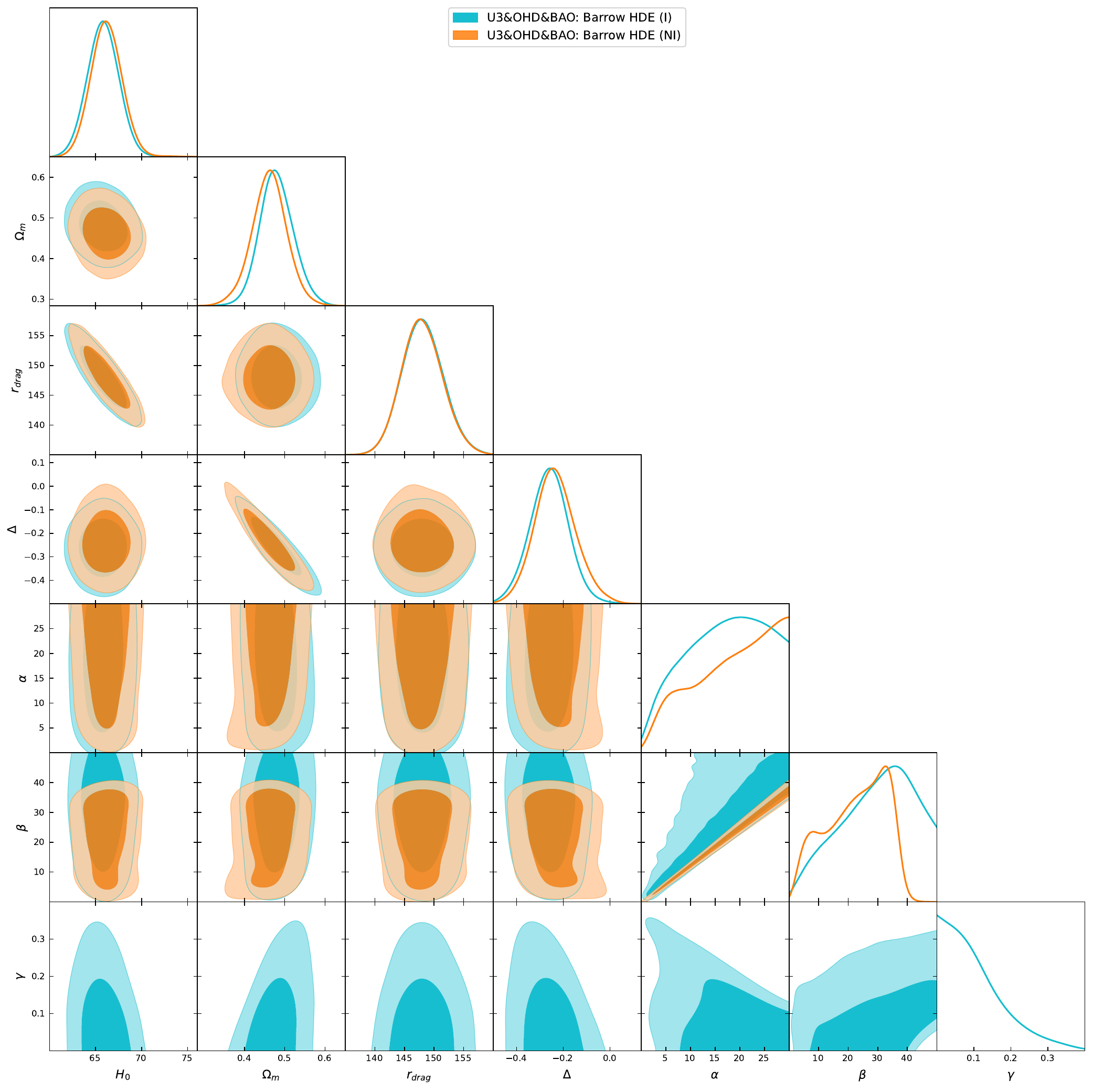}\caption{Confidence
space for the best-fit parameters for the interacting BHDE model.}%
\label{fig2}%
\end{figure}

Moreover, from Fig.~\ref{fig1} we observe a correlation between the parameters $\alpha$ and $\beta$, which is present in both the non-interacting and interacting scenarios. This degeneracy is particularly evident when the U3 dataset is employed, whereas a mild tension between the two parameters emerges in the PP-based analysis. This behavior can be traced back to the different sensitivity of the two supernova compilations to late-time dark energy dynamics. Indeed, it is well known that the Union3 catalogue exhibits a stronger preference for a dynamical dark energy component compared to PantheonPlus, which in turn allows for a wider region of the $(\alpha,\beta)$ parameter space to be observationally viable. From a theoretical standpoint, the $\alpha$-$\beta$ degeneracy originates from the structure of the evolution equations \eqref{OMDE} and \eqref{intmo}, where the cosmological dynamics depends primarily on specific combinations of these parameters, such as the ratio $\alpha/\beta$, rather than on their individual values. As a result, variations in $\alpha$ can be efficiently compensated by corresponding changes in $\beta$, leading to elongated confidence regions and, in the case of the U3\&OHD\&BAO dataset, to weak or absent upper bounds on these parameters. This highlights the intrinsic limitation of background data alone in fully breaking the degeneracy associated with the GO cutoff.


\begin{table}[t] \centering
\begin{tabular}
[c]{cccccccccc}\hline\hline
& $\mathbf{H}_{0}$ & $\mathbf{\Omega}_{m0}$ & $\mathbf{r}_{drag}$ & $\Delta$ &
$\alpha$ & $\mathbf{\beta}$ & $\gamma$ & $\mathbf{\chi}_{\min}^{2}%
\mathbf{-\chi}_{\Lambda\min}^{2}$ & $\mathbf{AIC-AIC}_{\Lambda}$\\
\multicolumn{10}{c}{\textbf{Dataset}: $\mathbf{PP\&OHD\&BAO}$}\\
\textbf{BHDE (NI)} & $67.9_{-1.7}^{+1.7}$ & $0.358_{-0.02}^{+0.16}$ &
$147.5_{-3.5}^{+3.5}$ & $-0.041_{-0.34}^{+0.01}$ & $1.58_{-0.55}^{+0.23}$ &
$1.49_{-0.93}^{+0.49}$ & $-$ & $-4.27$ & $+1.73$\\
\textbf{BHDE (I)} & $67.9_{-1.7}^{+1.7}$ & $0.418_{-0.029}^{+0.079}$ &
$147.5_{-3.4}^{+3.4}$ & $-0.197_{-0.16}^{+0.06}$ & $1.20_{-0.56}^{+0.25}$ &
$1.31_{-0.78}^{+0.29}$ & $<0.086$ & $-4.27$ & $+3.73$\\
$\Lambda$\textbf{CDM} & $68.5_{-1.6}^{+1.6}$ & $0.31_{-0.012}^{+0.011}$ &
$147.1_{-3.4}^{+3.4}$ & $-$ & $-$ & $-$ & $-$ & $0$ & $0$\\
\multicolumn{10}{c}{\textbf{Dataset}: $\mathbf{U3\&OHD\&BAO}$}\\
\textbf{BHDE (NI)} & $66.2_{-1.7}^{+1.7}$ & $0.462_{-0.044}^{+0.044}$ &
$148.0_{-3.5}^{+3.5}$ & $-0.234_{-0.093}^{+0.082}$ & $>14.6$ & $23_{-7}^{+10}$
& $-$ & $-9.63$ & $-3.63$\\
\textbf{BHDE (I)} & $65.8_{-1.7}^{+1.7}$ & $0.480_{-0.044}^{+0.039}$ &
$148.1_{-3.5}^{+3.5}$ & $-0.260_{-0.083}^{+0.083}$ & $17.0_{-5.0}^{+12}$ &
$30_{-8}^{+20}$ & $<0.127$ & $-9.67$ & $-1.67$\\
$\Lambda$\textbf{CDM} & $68.6_{-1.6}^{+1.6}$ & $0.31_{-0.014}^{+0.014}$ &
$146.7_{-3.4}^{+3.4}$ & $-$ & $-$ & $-$ & $-$ & $0$ & $0$\\\hline\hline
\end{tabular}
\caption{Observational constraints on the BHDE and $\Lambda$CDM models.}%
\label{bestfit}%
\end{table}%


\subsection{Evolution of the cosmological
parameters}

It is also interesting to study the
evolutions of the cosmological parameters such as $\Omega_{de}$,
$w_{de}$, $q$ and $v_s^2$ as functions of redshift $z$ for the
allowed ranges of the model parameters derived in the previous
subsection using data fitting. Let us consider the noninteracting
and interacting cases, separately. Note that $\delta$ in these
figures is the Barrow exponent, namely $\delta=\Delta$.

\subsubsection{Noninteracting case}

\begin{figure*}[!ht]
\centering
\includegraphics[trim={0cm 0cm 0cm 0.27cm},clip,scale=0.65]{{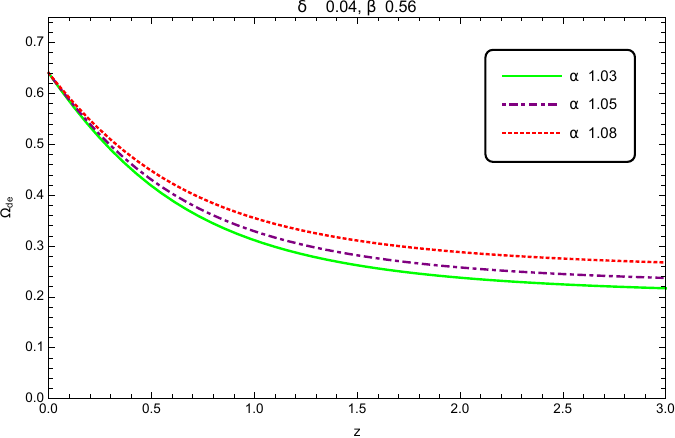}}
\hspace{5mm}
\includegraphics[trim={0cm 0cm 0cm 0.31cm},clip,scale=0.65]{{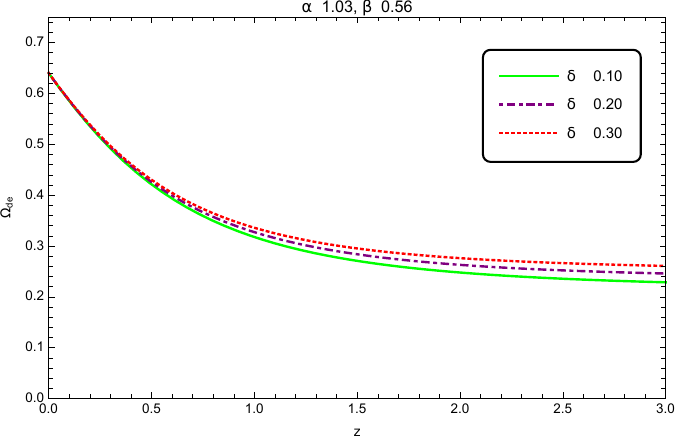}}
\caption{Evolution of $\Omega_{de}$ against $z$ for
BHDE in a flat cosmology for $\Omega_{m,0}=0.36$ and $\beta=0.56$.
In the left panel $\delta=-0.04$, while in the right panel
$\alpha=1.03$.}
 \label{Fig1}
\end{figure*}

\begin{figure*}[!ht]
\centering
\includegraphics[trim={0cm 0cm 0cm 0.27cm},clip,scale=0.65]{{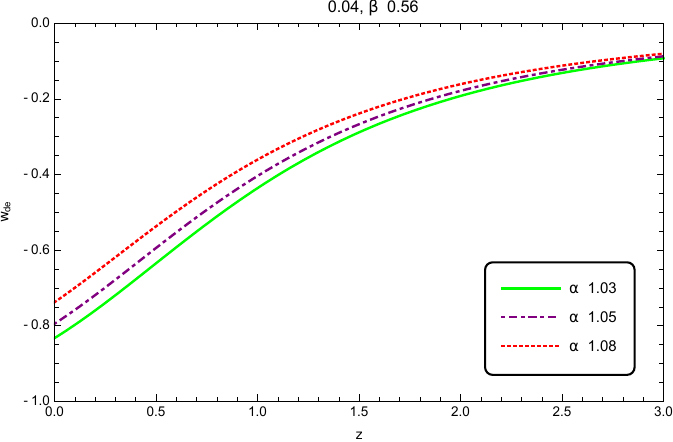}}
\hspace{5mm}
\includegraphics[trim={0cm 0cm 0cm 0.31cm},clip,scale=0.65]{{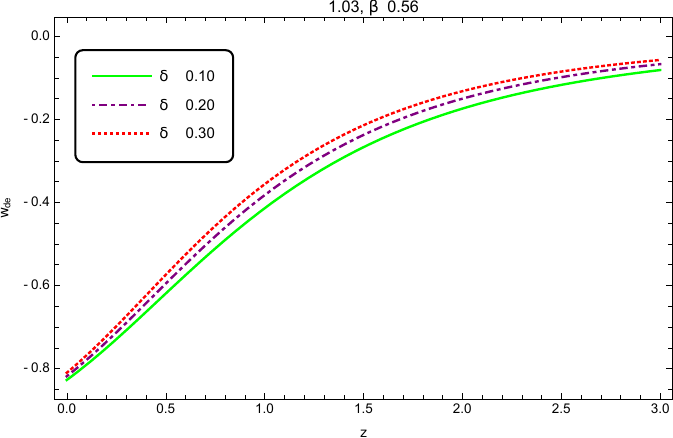}}
\caption{ Evolution of $w_{de}$ against $z$ for BHDE in
a flat cosmology for $\Omega_{m,0}=0.36$ and $\beta=0.56$. In the
left panel $\delta=-0.04$, while in the right panel
$\alpha=1.03$.}
 \label{Fig2}
\end{figure*}

\begin{figure*}[!ht]
\centering
\includegraphics[trim={0cm 0cm 0cm 0.27cm},clip,scale=0.65]{{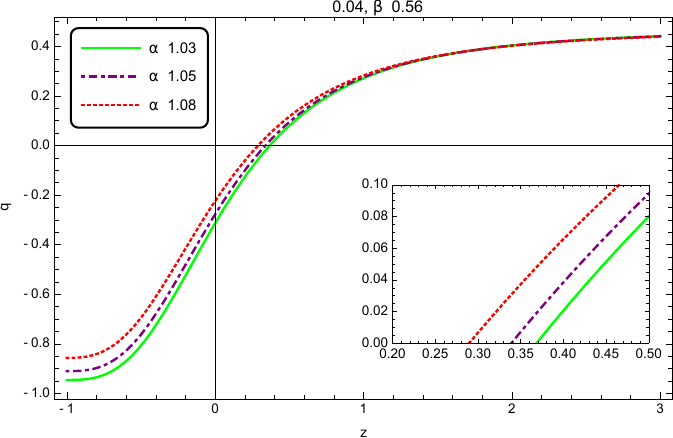}}
\hspace{5mm}
\includegraphics[trim={0cm 0cm 0cm 0.31cm},clip,scale=0.65]{{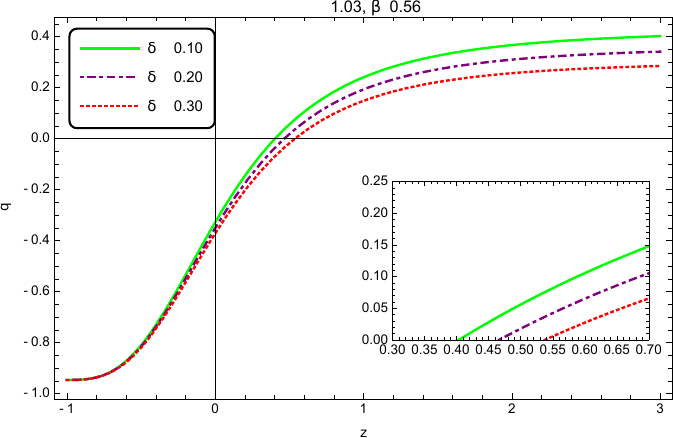}}
\caption{Evolution of the deceleration parameter $q$
against $z$ for BHDE in a flat cosmology for $\Omega_{m,0}=0.36$
and $\beta=0.56$. In the left panel $\delta=-0.04$, while in the
right panel $\alpha=1.03$.}
 \label{Fig3}
\end{figure*}

For the noninteracting case, the evolutions of
$\Omega_{de}$, $w_{de}$, $q$  are governed by Eqs. (\ref{OMDE}),
(\ref{wd}) and (\ref{q1}). In Figs. (\ref{Fig1})-(\ref{Fig3}), we
have plotted the behaviour of these parameters against $z$. From
these figures, we observe that the behaviour of the cosmological
parameters are compatible with recent observation for the allowed
ranges of $\alpha$, $\beta$ and $\delta$. Here we have taken the
values of these parameters in the ranges given by data fitting.

\subsubsection{Interacting case}

\begin{figure*}[!ht]
\centering
\includegraphics[trim={0cm 0cm 0cm 0.27cm},clip,scale=0.65]{{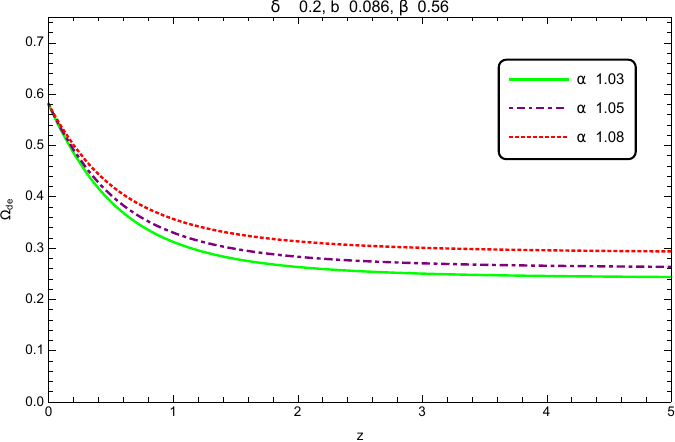}}
\hspace{5mm}
\includegraphics[trim={0cm 0cm 0cm 0.31cm},clip,scale=0.65]{{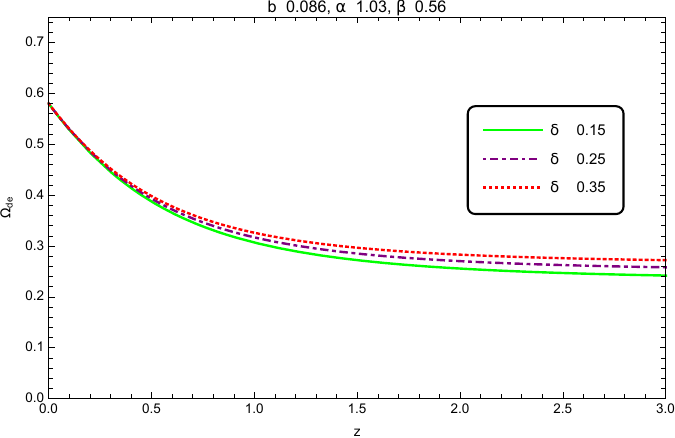}}
\caption{Evolution of $\Omega_{de}$ against $z$ for
interacting BHDE in a flat cosmology for $\Omega_{m,0}=0.42$,
$\gamma=0.086$ and $\beta=0.56$. In the left panel $\delta=-0.2$,
while in the right panel $\alpha=1.03$.}
 \label{figin1}
\end{figure*}

\begin{figure*}[!ht]
\centering
\includegraphics[trim={0cm 0cm 0cm 0.27cm},clip,scale=0.65]{{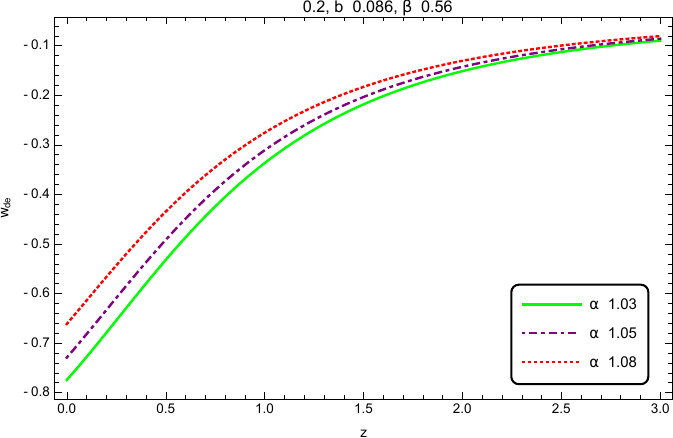}}
\hspace{5mm}
\includegraphics[trim={0cm 0cm 0cm 0.31cm},clip,scale=0.65]{{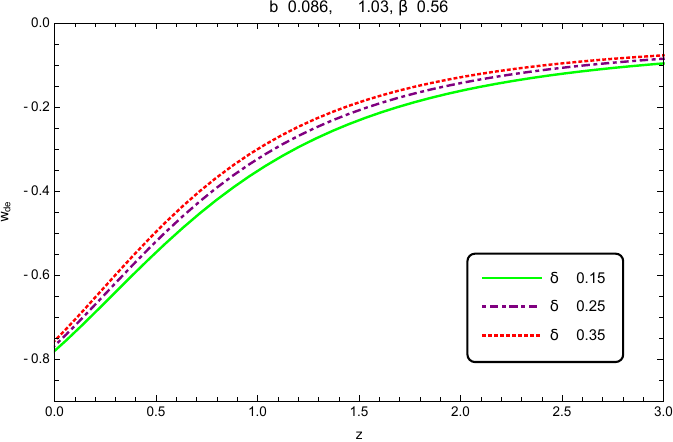}}
\caption{Evolution of $w_{de}$ against $z$ for
interacting BHDE in a flat cosmology for $\Omega_{m,0}=0.42$,
$\gamma=0.086$ and $\beta=0.56$. In the left panel $\delta=-0.2$,
while in the right panel $\alpha=1.03$.}
 \label{figin2}
\end{figure*}

\begin{figure*}[!ht]
\centering
\includegraphics[trim={0cm 0cm 0cm 0.27cm},clip,scale=0.65]{{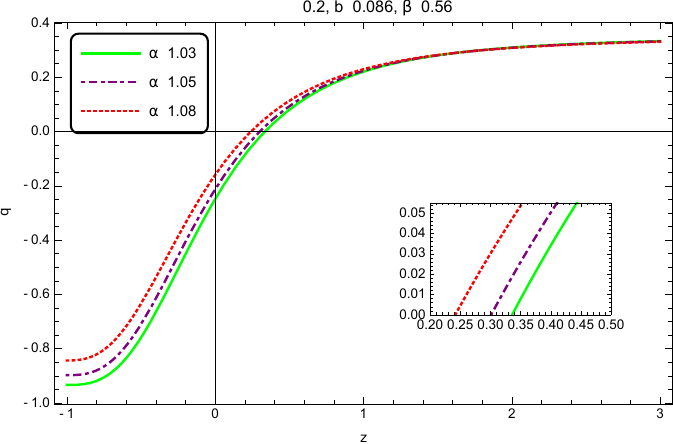}}
\hspace{5mm}
\includegraphics[trim={0cm 0cm 0cm 0.31cm},clip,scale=0.65]{{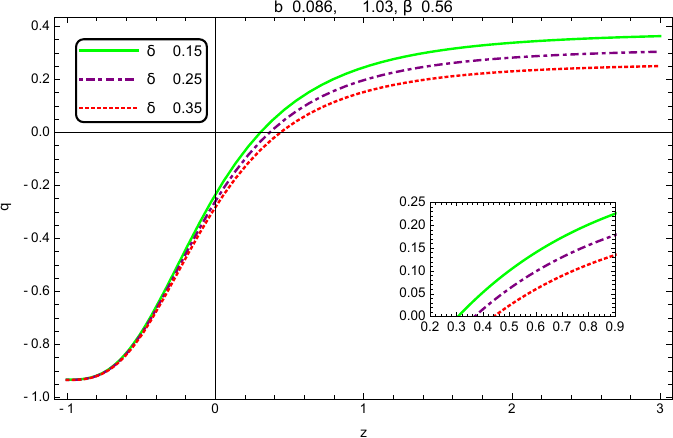}}
\caption{Evolution of the deceleration parameter $q$
against $z$ for interacting BHDE in a flat cosmology for
$\Omega_{m,0}=0.42$, $\gamma=0.086$ and $\beta=0.56$. In the left
panel $\delta=-0.2$, while in the right panel $\alpha=1.03$.}
 \label{figin3}
\end{figure*}

For the interacting case, the  dynamics of
$\Omega_{de}$ is followed by Eq. (\ref{master2}), while the
evolutions of $w_{de}$ and $q$ are still governed by Eqs.
(\ref{wd}) and (\ref{q1}). In Figs. (\ref{figin1})-(\ref{figin3}),
we have plotted the behaviour of these parameters against $z$.
Again, we see that the behaviour of the parameters are compatible
with recent observations.

Finally, we plot the evolution of squared sound
speed ($v^{2}_s$) to verify the instability of the model through a
semi-Newtonian approach. Here we only present two figures for some
allowed ranges of  the model parameters. The details of this study
can be found in \cite{Motaghi:2024rag}.

\begin{figure*}[!ht]
\centering
\includegraphics[trim={0cm 0cm 0cm 0.27cm},clip,scale=0.65]{{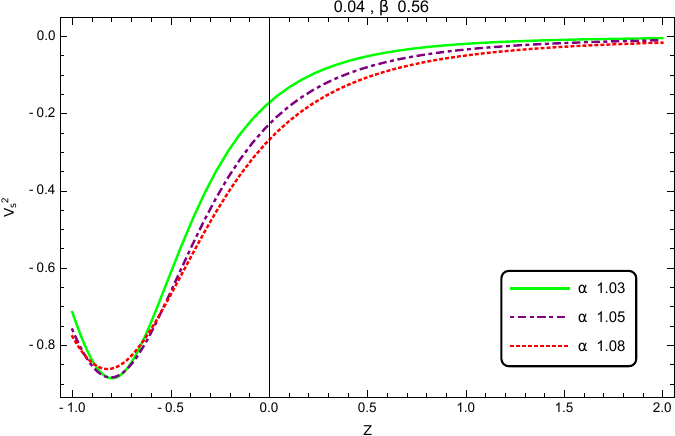}}
\hspace{5mm}
\includegraphics[trim={0cm 0cm 0cm 0.31cm},clip,scale=0.65]{{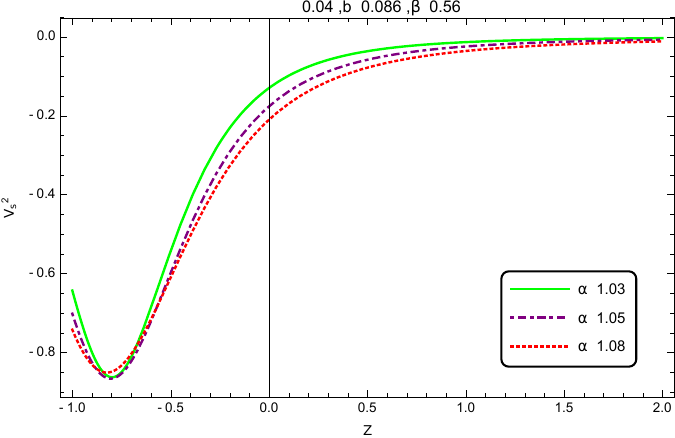}}
\caption{Evolution of $v_s^2$ against $z$ for BHDE in
a flat cosmology for $\Omega_{m,0}=0.36$, $\delta=-0.04$ and
$\beta=0.56$. Left panel for noninteracting case while right panel
for interacting case with $\gamma=0.086$.}
 \label{figin4}
\end{figure*}

\section{Conclusions}\label{sec5}
In this work, we have explored the cosmological consequences of
Barrow and Tsallis HDE implemented with the GO-IR cutoff. By
deriving the modified Friedmann equations that follow from the
Barrow entropy deformation and analyzing both non-interacting and
interacting dark sector configurations, we have investigated how
the entropic parameter $\Delta$ influences the late-time expansion
history. Confronting the theoretical predictions with several
state-of-the-art observational datasets-including Type Ia
supernovae from PantheonPlus and Union3, Hubble parameter
measurements from Cosmic Chronometers and BAO distance indicators
from DESI DR2-we obtained tight constraints on the BHDE parameter
space through a Bayesian MCMC analysis.

Our results indicate that the BHDE scenario provides a robust
description of the late-time Universe and remains fully compatible
with current observations. Overall, the model provides a fit to
the data comparable to, and in some cases slightly better than,
that of $\Lambda$CDM, although the concordance model still retains
a marginal statistical preference according to the Akaike
Information Criterion once differences in model complexity are
taken into account. This trend is, however, reversed when the
U3\&OHD\&BAO dataset is considered. In this case the
non-interacting BHDE scenario shows a weak statistical preference
over $\Lambda$CDM, while the interacting version becomes
statistically equivalent to the concordance model.

Notably, our analysis favors slightly negative values of the
Barrow entropic index $\Delta$, in agreement with recent
independent studies and with the theoretical allowance for
anomalous fractal deformations of the horizon geometry
\cite{Tang,Dagotto:1989gp,Luciano:2025elo,Luciano:2025hjn}. These
findings reinforce the relevance of generalized entropy frameworks
as promising extensions of HDE, offering a richer phenomenology
while preserving consistency with late-time cosmological probes. It is worth stressing that, although our analysis has been carried out explicitly within the Barrow entropy framework, the results straightforwardly extend to the Tsallis HDE scenario. Indeed, as emphasized in the Introduction, Barrow and Tsallis entropies share the same functional dependence on the horizon area, provided the correspondence $\Delta \rightarrow 2(\epsilon-1)$ is applied. Consequently, the cosmological dynamics and observational constraints derived in this work are equally valid for Tsallis HDE upon a simple reparametrization of the entropic index. In this light, the preference for slightly negative values of the Barrow parameter $\Delta$ translates into $\epsilon<1$ in the Tsallis formulation, corresponding to a sub-extensive scaling of the Tsallis entropy. This regime is well motivated within non-extensive statistical mechanics and further supports the physical viability of generalized entropy-based holographic dark energy models (see, e.g. \cite{Luciano:2022ely} for a review of recent cosmological bounds on Tsallis parameter).

Several avenues merit further exploration. A natural extension of
this work is the study of BHDE at the perturbative level, focusing
on structure growth and, in particular, the $\sigma_8$ tension.
Such analyses would allow one to assess whether entropic
modifications can simultaneously describe both the expansion
history and the clustering properties of matter. Incorporating CMB
temperature and polarization spectra would further strengthen the
constraints and test the consistency of the model across cosmic
epochs. Additionally, the framework provides an intriguing context
for examining the Hubble tension. Indeed, generalized entropy
corrections may introduce subtle modifications to the late-time
expansion rate, and future work could investigate whether BHDE can
offer a physically motivated route toward reconciling
high-redshift and low-redshift determinations of $H_0$.

Taken together, our results highlight Barrow and Tsallis HDE
models with the GO cutoff as viable and competitive extension of
the standard dark energy paradigm. A comprehensive investigation
of its perturbative predictions, early-Universe imprints and
implications for cosmological tensions will be essential for fully
assessing the role of generalized entropy in the gravitational and
cosmological sectors. These directions will be the subject of
future work.
\begin{acknowledgments}
The research of GGL is supported by the postdoctoral fellowship
program of the University of Lleida. GGL gratefully acknowledges
the contribution of the LISA Cosmology Working Group (CosWG), as
well as support from the COST Actions CA21136 - \textit{Addressing
observational tensions in cosmology with systematics and
fundamental physics (CosmoVerse)} - CA23130, \textit{Bridging high
and low energies in search of quantum gravity (BridgeQG)} and
CA21106 - \textit{COSMIC WISPers in the Dark Universe: Theory,
astrophysics and experiments (CosmicWISPers)}. AP \& GL thank the
support of VRIDT through Resoluci\'{o}n VRIDT No. 096/2022 and
Resoluci\'{o}n VRIDT No. 098/2022. Part of this study was
supported by FONDECYT 1240514.
\end{acknowledgments}

\bibliography{Bib}
\end{document}